\documentclass{ws-procs9x6-cpt16}
\begin{document}

\newcommand{\refeq}[1]{(\ref{#1})}
\def\etal {{\it et al.}}

\title{The Elusive Part of the Standard-Model Extension Gravitational Sector}

\author{Yuri Bonder}

\address{Instituto de Ciencias Nucleares, Universidad Nacional Aut\'onoma de M\'exico\\ Apartado Postal 70-543, Coyoac\'an, 04510, Cd. Mx., M\'exico
}

\begin{abstract}
In the minimal gravitational sector of the Standard-Model Extension, there is a coefficient whose physical consequences are  unknown, and the reason behind this lack of effects is still puzzling. This contribution summarizes several studies where the goal was to find a fundamental explanation of this puzzle. So far, no evidence of such a fundamental explanation has been found, suggesting that this coefficient could actually produce physical effects. Nevertheless, while looking for this fundamental reason, several relevant lessons have been revealed. 
\end{abstract}

\bodymatter

\section{Motivation}

The minimal gravitational sector of the Standard-Model Extension (mgSME) is described by the action for conventional physics plus the Lorentz violating term
\begin{equation}
S_{\rm mgSME} = \frac{1}{2\kappa}\int d^4 x \sqrt{-g}k^{\mu\nu\rho\sigma} R_{\mu\nu\rho\sigma},
\end{equation}
where $\kappa$ is the coupling constant of general relativity (GR), $g$ is the determinant of the metric $g_{\mu\nu}$, $k^{\mu\nu\rho\sigma}$ are the Lorentz-violation (LV) coefficients, and $R_{\mu\nu\rho\sigma}$ is the Riemann tensor. Note that, since this work concerns curved spacetimes, LV has to arise spontaneously,\cite{Kostelecky2004} and thus $k^{\mu\nu\rho\sigma}$ are dynamical. The $k^{\mu\nu\rho\sigma}$ can be separated into irreducible pieces as
\begin{equation}
k^{\mu\nu\rho\sigma} R_{\mu\nu\rho\sigma}=-uR + s^{\mu\nu} R^T_{\mu\nu} +t^{\mu\nu\rho\sigma}W_{\mu\nu\rho\sigma},
\end{equation}
where $R$, $R^{T}_{\mu\nu}$, and $W_{\mu\nu\rho\sigma}$ stand, respectively, for the curvature scalar, the traceless Ricci tensor, and the Weyl tensor. Note that $s^{\mu\nu}$ and $t^{\mu\nu\rho\sigma}$ share the index symmetries of $R^T_{\mu\nu}$ and $W_{\mu\nu\rho\sigma}$, respectively.

Remarkably, to date, the effects of the $t^{\mu\nu\rho\sigma}$ coefficient are still unknown; this is known as the $t$ puzzle.\cite{BaileyKostelecky,RuiQuentinAlan} This contribution is devoted to describing several analyses where a fundamental explanation for this puzzle is sought.

\section{Field redefinitions}

It is well known\cite{field redef bibliog} that field redefinitions can be used to move some LV coefficients to other sectors of the Standard-Model Extension (SME). It is thus natural to expect that a field redefinition could explain the $t$ puzzle. In GR, the metric is the dynamical field. Therefore, it is tempting to study what LV coefficients arise, in the GR action $S_{\rm EH}$, with a redefinition $g_{\mu\nu}\rightarrow \tilde{g}_{\mu\nu}$. With a particular metric redefinition, it is possible to show\cite{tpuzzle} that, to first order in the LV coefficients,
\begin{equation}
S_{\rm EH} \rightarrow S_{\rm EH} + \frac{1}{2\kappa} \int d^4x \sqrt{-\tilde{g}} \left[-uR(\tilde{g}) + s^{\mu\nu}R^T_{\mu\nu}(\tilde{g})\right],
\end{equation}
where a total divergence, which has no physical effects, has been ignored. This result implies that the $u$ and $s^{\mu\nu}$ coefficients can be moved to other SME sectors, which is consistent with previous results in linearized metric approximation.\cite{KosteleckyTasson} In addition, it proves that $t^{\mu\nu\rho\sigma}$ cannot be removed with a metric redefinition.

In GR there are two equivalent dynamical formalisms: the standard and the Palatini. In the first approach, the metric is the only dynamical field and the (torsionless) connection is determined by requiring that the covariant derivative of the metric vanishes. In the latter approach, the metric and the connection are assumed to be dynamically independent fields and the equation of motion for the connection yields the condition that the metric covariant derivative vanishes.\cite{Palatini} If the metric and the connection can be treated as independent fields, it is possible to perform more general field redefinitions. Moreover, it has been shown that, to first order in the LV coefficients, the mgSME yields the same physical predictions in both approaches.\cite{tpuzzle} However, these independent redefinitions are not extremely revealing: the metric redefinition leads to the $u$ and $s^{\mu\nu}$ terms (with no divergence), while the connection redefinition produces new terms that are definitively not of the form of $t^{\mu\nu\rho\sigma}W_{\mu\nu\rho\sigma}$.\cite{tpuzzle} Therefore, the $t^{\mu\nu\rho\sigma}$ term cannot be removed with field redefinitions of the gravitational fields.

\section{Lanczos-like tensor}

An analytic tensor with the index symmetries of the Weyl tensor can be written in terms of the covariant derivative ($D_\mu$) of a `Lanczos potential' $H^{\mu\nu\rho}$.\cite{Bampi} This potential is such that $H^{\mu\nu\rho} =- H^{\nu\mu\rho}$, and $H^{[\mu\nu\rho]}$, $g_{\nu\rho}H^{\mu\nu\rho}$, and $D_\rho H^{\mu\nu\rho}$ vanish. After replacing $t^{\mu\nu\rho\sigma}$ by its Lanczos potential, the mgSME action takes the form
\begin{equation}\label{mgSME action Lanczos}
S_{\rm mgSME} = \frac{1}{2\kappa}\int d^4 x \sqrt{-g} \left[ -u R + s^{\mu\nu} R^T_{\mu\nu} +4 H^{\mu\nu\rho}D_\mu R_{\nu\rho}\right].
\end{equation}
Observe that the $t^{\mu\nu\rho\sigma}$ term has been converted into a dimension-$5$ operator, and this type of operator is known to generate, in the nonrelativistic weak-gravity approximation, unphysical self accelerations.\cite{RuiQuentinAlan} This may be the reason behind the $t$ puzzle, but it is not a fundamental explanation.

It is also tempting to integrate the last term in Eq.~\refeq{mgSME action Lanczos} by parts, obtaining an effective $s^{\mu\nu}$ coefficient: $s_{\rm eff}^{\mu\nu} = s^{\mu\nu}- 4 D_\rho H^{\rho \mu\nu}$. However, $s_{\rm eff}^{\mu\nu}$ depends on the metric (through the covariant derivative), and thus it cannot be considered as an LV coefficient. Still, it should be stressed that, in the linearized gravity approximation and neglecting terms proportional to the LV coefficients and the metric perturbation, this procedure accounts for the absence of physical effects associated with $t^{\mu\nu\rho\sigma}$.

\section{Other ideas}

It is well known that, if the spacetime under consideration has boundaries, $S_{\rm EH}$ needs to be corrected with the so-called York--Gibbons--Hawking boundary term to lead to Einstein's equations.\cite{HG term} This could be relevant for the $t$ puzzle since, typically, the phenomenological studies in the SME involve conformally flat spacetimes, which have boundaries. Therefore, it should be verified if a boundary term can be constructed for all the coefficients in the mgSME. It turns out that such a boundary term exists for all coefficients in the mgSME, including $t^{\mu\nu\rho\sigma}$.\cite{tpuzzle} However, such a term cannot be constructed in the nonminimal sector, which needs to be carefully handled in the presence of spacetime boundaries.

The last idea is related to the Cauchy problem\cite{Wald} for the action of the LV coefficients. Generically, it is hard to study this problem. Therefore, it is useful to focus on a simpler case. In a particular example of the so-called bumblebee models\cite{Bumblebee} where the vector field has a Maxwell kinetic term and a potential that drives the spontaneous Lorentz violation, it has been shown that there exists a Hamilton density that generates a constraint-compatible evolution, but that the evolution is not uniquely determined by the required initial data.\cite{BonderEscobar} In the future, the question that needs to be analyzed is whether the feasible actions for $t^{\mu\nu\rho\sigma}$ have well-posed Cauchy problems.

\section{Conclusions}

To date, the physical effects of a coefficient in the mgSME, $t^{\mu\nu\rho\sigma}$, remain unknown. While searching for a fundamental reason for this puzzle, some lessons were learned: (1) redefinitions of the gravitational fields generate $u$ and $s^{\mu\nu}$ but no $t^{\mu\nu\rho\sigma}$, (2) the mgSME can be treated \textit{\`a la Palatini}, (3) it is possible to correct the mgSME action to cancel spacetime boundary effects, but there is no boundary term for the nonminimal sector, and (4) the Cauchy problem for theories with spontaneous Lorentz violation may be ill posed.

The fact that no fundamental explanation for the $t$ puzzle has been found suggests that $t^{\mu\nu\rho\sigma}$ could be physical, and that the phenomenological methods that have been used could be hiding its effects. It thus seems promising to look for the effects of $t^{\mu\nu\rho\sigma}$ in different phenomenological schemes.

\section*{Acknowledgments}
This work was done with financial support from UNAM-DGAPA-PAPIIT Project No. IA101116.


\begin{thebibliography}{xx}
\bibitem{Kostelecky2004}
V.A.\ Kosteleck\'y, 
Phys.\ Rev.\ D {\bf 69}, 105009 (2004).

\bibitem{BaileyKostelecky}
Q.G.\ Bailey and V.A.\ Kosteleck\'y, 
Phys.\ Rev.\ D {\bf 74}, 045001 (2006); 
B.~ Altschul, Q.G.\ Bailey, and V.A.\ Kosteleck\'y, 
Phys.\ Rev.\ D {\bf 81}, 065028 (2010).

\bibitem{RuiQuentinAlan}
Q.G.\ Bailey, V.A.\ Kosteleck\'y, and R.\ Xu, 
Phys.\ Rev.\ D {\bf 91}, 022006 (2015).

\bibitem{field redef bibliog}
V.A.\ Kosteleck\'y and M.\ Mewes, 
Phys.\ Rev.\ D {\bf 66}, 056005 (2002);
R.\ Lehnert, Phys.\ Rev.\ D {\bf 74}, 125001 (2006);
Y.\ Bonder, Phys.\ Rev.\ D {\bf 88}, 105011 (2013).

\bibitem{tpuzzle}
Y.\ Bonder, Phys.\ Rev.\ D {\bf 91}, 125002 (2015).

\bibitem{KosteleckyTasson}
V.A.\ Kosteleck\'y and J.D.\ Tasson, 
Phys.\ Rev.\ D {\bf 83}, 016013 (2011).

\bibitem{Palatini}
M.\ Ferraris, M.\ Francaviglia, and C.\ Reina, 
Gen.\ Rel.\ Grav.\ {\bf 14}, 243 (1982).

\bibitem{Bampi}
F.\ Bampi and G.\ Caviglia, 
Gen.\ Rel.\ Grav.\ {\bf 15}, 375 (1983).

\bibitem{HG term}
J.W.\ York, 
Phys.\ Rev.\ Lett.\ {\bf 28}, 1082 (1972);
G.W.\ Gibbons and S.W.\ Hawking, 
Phys.\ Rev.\ D {\bf 15}, 2752 (1977).

\bibitem{Wald}
R.M.\ Wald, 
{\it General Relativity}, University of Chicago Press, 
Chicago, 1984, chapter 10.

\bibitem{Bumblebee}
R.\ Bluhm and V.A.\ Kosteleck\'y,
Phys.\ Rev.\ D {\bf 71}, 065008 (2005).

\bibitem{BonderEscobar}
Y.\ Bonder and C.A.\ Escobar, 
Phys.\ Rev.\ D {\bf 93}, 025020 (2016).

\end{thebibliography}
\end{document}